\documentstyle[12pt]{article}
\input tcilatex
\begin{document}

\begin{center}
{\bf Static spherically symmetric constant density relativistic and
Newtonian stars in the Lobachevskyan geometry.}

\vspace*{1.5cm}

S.M.KOZYREV

Scientific center gravity wave studies ''Dulkyn''.

e-mail: Sergey@tnpko.ru
\end{center}

\vspace*{1.5cm}

\begin{center}
${\bf Abstract}$
\end{center}

The present paper has the purpose to illustrate the importance of the ideas
and constructions of the Non-Euclidean (Lobachevsky) Geometry, which can be
applied even today for solving some conceptually important problems. We
study the static and spherically symmetric solutions to the Einstein field
equations under the assumption that the space-time may possess an arbitrary
number of spatial dimensions. A new exact solution of a perfect fluid sphere
of constant (homogeneous) energy-density which agrees with interior
Lobachevsky geometry for 3D and 4D spaces are found. We discuss the property
of temporal scalar field arise in lower-dimensional theories as the
reduction of extra dimension.

\section{Introduction}

Since the pioneering work of Lobachevsky \cite{1} the aim of solving the
problems on the astronomical verification of the geometry of our visible
world and on the kinds of changes which will occur in mechanics after
introducing in it the new geometry has been pursued. In the framework of
Lobachevsky hypothesis the forces produce all by themselves the motion,
velocity, time, mass and even distances and angle \cite{10} . The
lower-dimensional theories of gravity are a good candidate for analyze this
problems while is that it is simple enough to be soluble but yet contains
non-trivial features. The lower-dimensional theories of gravity have been
intensively investigated, primarily because of they can assume as a
theoretical laboratories for applying techniques that appear intractable or
awkward in the in four-dimensional general relativity. There are other
reasons for studying dynamical gravity in three dimensions as membrane
models of extended relativistic systems. The 3-membrane is a
three-dimensional immersion in some higher-dimensional space. Its dynamics
can be formulated in terms of a three-dimensional field theory and its
internal stresses described in terms of an intrinsic geometry induced by the
immersion.

It is surprising that in 2+1 dimensions, Einstein general theory of
relativity has very little in common with gravitation in a spacetime vacuum
having four dimensions. The Riemann-Christoffel tensor is uniquely
determined by the Ricci tensor, which vanishes outside the sources. Hence,
the gravitational dynamics of test particles must then be induced by
topological effects in flat space do not feel any gravitational field. Thus
vacuum Einstein gravity in three dimensions is devoid of any limit that may
be identified with an analogue of Newtonian gravity \cite{2}, \cite{20}. A
proper relativistic theory of gravitation in 3 dimensions needs some
additional ingredient besides the metric tensor. It is well known, in
dimension lower than four, this ingredient present in Jordan, Brans-Dicke
theory - the scalar field is essential to have a Newtonian regime at low
energies. Furthermore, the metric in 2+1 dimensions in which the curvature
tensor vanishes and gravity is due to torsion (teleparallel theory) gives
the correct Newtonian limit \cite{3}.

A simple and elegant idea in this connection employ the equivalence between
the (D + 1) Kaluza-Klein theories with D-dimensional Einstein theory with
source \cite{c31}, \cite{c32}. Furthermore, this correspondence gives a
mechanism for formulation of a 3-dimensional theory containing temporal
scalar field. In this scheme space and time are treated in completely
different ways. This might seem to be a retrogressive step, but actually the
approach achieves everything that Einstein did by presupposing a
four-dimensional unity of space and time.

This paper details the new approach to the problem of separate the temporal
scalar field from the four-dimensional metric context of general relativity
which where outlined in previous letter \cite{c33}. Section 2 develops the
formalism related to the correspondence between the ordinary 4 - dimensional
gravity with 3 - dimensional theories with temporal scalar field. In Section
3 the exact solutions for the static constant density homogeneous sphere has
been obtained. Finally, Sec. 4 briefly discussed some aspects of these
theories. We choose units such that G$_N$ = c = 1, and let Latin indices run
0-3 and Greek indices run 0-2.

\section{ Equivalence between Einstein and 3 dimensional theories with
temporal scalar field.}

The investigation of a lower dimension theory of gravity with additional
scalar field appears to be the way to construct a model with properties more
akin to those of general relativity. The most natural way to introduce the
scalar field is a setting up a correspondence between the ordinary 4 -
dimensional gravity with 3 - dimensional theories with source. We do not
claim that this is the only possibility but this would bring some more
clarification on the role of the scalar fields in physics. The procedure to
go from D+1 to D is by now well known (see \cite{c31}, \cite{c32}). Consider
the case of 4 dimensional vacuum field equations

\begin{center}
\begin{equation}
^4R_{ab}=0,  \label{1.1}
\end{equation}
\end{center}

where R$_{ab}$ is a Ricci tensor in the 4 - dimensional space. It is well
known that equation (\ref{1.1}) gives rise to 3 dimensional Einstein
equations with source in the form \cite{c32}:\thinspace 
\begin{eqnarray}
^3R_{\alpha \beta }=^3T_{\alpha \beta }^\phi ,  \label{1.2}
\end{eqnarray}

where $T_{\alpha \beta }^\phi $ is the induced energy-momentum tensor.

Following earlier work, it is inquired how far the 4 - dimensional equations
without sources may be reduced to the Einstein equations with sources. It is
shown by algebraic means that this can be done, provided the extra part of
the 4 - dimensional geometry is used appropriately to define an effective
3-D energy-momentum tensor of a scalar field. We start with the 4
dimensional source-free Einstein field equations (\ref{1.1}) and take the
metric to be in the form:

\begin{equation}
^4g_{ab}=\left( 
\begin{tabular}{ll}
$^3g_{\alpha \beta }$ & 0 \\ 
0 & $g_{dd}$%
\end{tabular}
\right) ,  \label{1.3}
\end{equation}

where both the 4 and 3 dimensional metrics g$_{ab}$ and g$_{a\ss }$ are in
general dependent on the reducible coordinate x$^d$ and we write the g$_{dd}$
as

\begin{equation}
g_{dd}=\phi ^2,g^{dd}=\frac 1{\phi ^2}.  \label{1.4}
\end{equation}

Now the source free field equations in 4 dimensions are given by

\begin{eqnarray}
\begin{array}{cc}
& ^4R_{\alpha \beta }=0 \\ 
^4R_{ab}=0\Rightarrow & ^4R_{dd}=0 \\ 
& ^4R_{\alpha d}=0
\end{array}
\label{2.2}
\end{eqnarray}

and

\begin{equation}
\begin{array}{l}
^4R_{\alpha \beta }=\,^{\,3}R_{\alpha \beta }- \\ 
\,\,\,\,\,\,\,\,\,\,\,\,\,\,\,\,\,\,\,\,\,\,\,\,\,\,\,-\frac{\phi _{\alpha
;\beta }}\phi \frac 1{2\phi ^2}\left( \frac{\phi ,_d}\phi g_{\alpha \beta
,d}-g_{\alpha \beta ,dd}+g^{\lambda \mu }g_{\alpha \lambda ,d}g_{\beta \mu
,d}-\frac 12g^{\mu \nu }g_{\mu \nu ,d}g_{\alpha \beta ,d}\right) ,
\end{array}
\end{equation}

In this way the 4-dimensional source free Einstein type equations are
related to a 3 dimensional theory with sources, from the (\ref{2.2}) we
obtain

\begin{equation}
^3R_{\alpha \beta }-\frac 12~\,^3Rg_{\alpha \beta }=~^3T_{\alpha \beta
,}^\phi  \label{1.5}
\end{equation}

\begin{equation}
\ \Box \phi =-\frac 1{4\phi }g^{\lambda \mu }\,_{,d\,}g_{\lambda \nu
,d}-\frac 1{2\phi }g^{\lambda \mu }g_{\lambda \mu ,dd}+\frac{\phi ,_d}{2\phi
^2}g^{\lambda \mu }g_{\lambda \mu ,d},  \label{1.7}
\end{equation}

\begin{equation}
\ \left[ \frac 1{2\sqrt{g_{dd}}}\left( g^{\beta \mu }g_{\mu \alpha
,d}-\delta _\alpha ^\beta \,g^{\rho \nu }g_{\rho \nu ,d}\right) \right]
_{;\beta }=0.  \label{1.8}
\end{equation}

These give the components of the induced energy-momentum tensor since
Einsteins equations (\ref{1.5}) hold.

\begin{equation}
\begin{array}{r}
^3T_{\alpha \beta }=\frac{\phi _{\alpha ;\beta }}\phi -\frac 1{2\phi
^2}\left[ \frac{\phi ,_d}\phi g_{\alpha \beta ,d}-g_{\alpha \beta
,dd}+g^{\lambda \mu }g_{\alpha \lambda ,d}g_{\beta \mu ,d}-\frac 12g^{\mu
\nu }g_{\mu \nu ,d}g_{\alpha \beta ,d}\right] + \\ 
+\frac{g_{\alpha \beta }}{8\phi ^2}\left[ g^{\mu \nu },_{d\,\,}g_{\mu \nu
,d}+\left( g^{\mu \nu }g_{\mu \nu ,d}\right) ^2\right] .
\end{array}
\label{1.6}
\end{equation}

The generally covariant d'Alembertian $\Box $ is defined to be covariant
divergence of $\phi ^{,\rho }$:

\[
\Box \phi =\phi ^{,\rho }\,_{;\rho }=\sqrt{-g}\left( \sqrt{-g}\phi ^{,\rho
}\right) _{,\rho }. 
\]

From the above it is seen that, the physical metric obtained in this way can
then be used to evaluate possible effects from the extra dimension. This
means that it can employ the equivalence between spacelike or timelike
dimensions with the scalar field, respectively. In the first case we find
the field equations are identical to the Jordan, Brans-Dicke field equations
in 2+1 dimensions with the free parameter $\omega $ = 0. However, timelike
extra dimensions reduction procedure from 4 dimensions leads to a spacetime
containing temporal scalar field. The addition of the temporal scalar field
to the Einstein tensor gives a unique theory

Note that in more general case stress-energy tensor of matter is assumed to
have been derived from the Lagrangian in the usual way.

\section{Spacetime geometry of static constant density homogeneous stars.}

The apparently simple problem of the static spherically symmetric perfect
fluid has by now generated hundreds of scientific papers. The first two
exact solution of Einsteins field equations were obtained by Schwarzschild,
soon after Einstein advent of general relativity. The first solution
describes the geometry of the space-time exterior to a prefect fluid sphere
in hydrostatic equilibrium. While the other, known as interior Schwarzschild
solution, corresponds to the interior geometry of a fluid sphere of constant
(homogeneous) energy-density. The importance of these two solutions is well
known. Granted, a completely general method to solve Einsteins field
equations does not exist and therefore in this paper we focus on spherically
symmetric system. To this end, let us consider the field produced by a
static and isotropic source in regions devoid of sources in D = 3. Taking
into account that the field equations of this theory was obtained using the
reduction procedure from usual 4-dimensional general relativity we can find
the corresponding solution of Einstein fields equations using the line
element in the form \cite{c4}:

\begin{eqnarray}
ds^2=e^\alpha dr^2+e^\beta \left( d\theta ^2+\sin ^2\theta \,d\varphi
^2\right) .  \label{e2.3}
\end{eqnarray}

where $\alpha $ and $\beta $ are functions of r alone. We assume that
spacetime is static; the metric and the scalar field can be chosen such that

\begin{eqnarray}
g_{a\ss ,t}\equiv \frac{\partial ~g_{a\ss }}{\partial ~t}=\phi _{,t}=0
\label{2.5}
\end{eqnarray}

The corresponding energy-momentum tensor of matter is specialized to that of
perfect fluid

\begin{eqnarray}
T_{\alpha \beta }^M=p~g_{\alpha \beta }+\left( p+\mu \right) U_\alpha U_\beta
\label{e2.6er}
\end{eqnarray}

where p and $\mu $ are the proper pressure and energy density, respectively.
U$_\alpha $ are the components of the fluid velocity, which verifies for the
static case g$_{\mu \nu }$ U$_\alpha $ U$_\beta $ = -1, it follows that

\begin{eqnarray}
T_{tt}^M=-\mu ~g_{tt},~T_{a\ss }^M\ =p~g_{a\ss }  \label{2.6_1}
\end{eqnarray}

Having added (\ref{2.3}) and (\ref{2.6_1}) into (\ref{1.5})and (\ref{1.7}),
we obtain the following equations for the metric (\ref{e2.3}):

\begin{eqnarray}
\begin{array}{c}
\frac 12\alpha ^{\prime }\beta ^{\prime }-\frac 12\beta ^{\prime ~2}+\frac{%
\alpha ^{\prime }\phi ^{\prime }}{2\phi }-\beta ^{\prime \prime }\frac{\phi
^{\prime \prime }}\phi =4e^\alpha \pi \left( p-\mu \right) \\ 
\\ 
e^{\alpha -\beta \,}+\frac 14\alpha ^{\prime }\beta ^{\prime }-\frac 12\beta
^{\prime ~2}-\frac{\beta ^{\prime \prime }}2-\frac{\beta ^{\prime }\,\phi
^{\prime }}{2\phi }=4e^\alpha \pi \left( p-\mu \right) \\ 
\\ 
-\left( \alpha ^{\prime }-2\beta ^{\prime }\right) \frac{\phi ^{\prime }}{%
2\phi }+\frac{2\,\phi ^{\prime \prime }}\phi =-4e^\alpha \pi \left( \mu
+3p\right)
\end{array}
\label{e2.7}
\end{eqnarray}

For the further specification of the problem an equation of state is needed.
We use the function of it a constant density. For this case the solutions of
equations (\ref{e2.7}) can be found analytically, by among the others there
is

\begin{eqnarray}
\begin{array}{c}
g=\left( 
\begin{array}{ccc}
1 & 0 & 0 \\ 
0 & \left( \kappa ~\sinh \left( \frac r\kappa \right) \right) ^2 & 0 \\ 
0 & 0 & \left( \kappa ~\sinh \left( \frac r\kappa \right) \sin \theta
\right) ^2
\end{array}
\right) \\ 
\\ 
\phi =c_2+\kappa ~c_1\cosh \left( \frac r\kappa \right) \\ 
\\ 
p=-\frac{c_2+3\kappa ~c_1\cosh \left( \frac r\kappa \right) }{8\pi ~\kappa
^2~\left( c_2+\kappa ~c_1\cosh \left( \frac r\kappa \right) \right) }
\end{array}
,  \label{e2.6}
\end{eqnarray}

where c$_1$ and c$_2$ arbitrary constant and $\kappa =\sqrt{\frac 3{8\pi
~\mu }}.~$We have seen that this solution provides the Lobachevsky geometry
in 3 dimensional spaces. The curvature of this space depends from the energy
density of matter. Thus in this sense the forth dimension generates
additional effective sources and these in turn act to curve the
3-dimensional spacetime too. Moreover the solution (\ref{e2.6}) should
correspond to an interior solution for the constant density stars of the
Einstein theory. In a suitable reference system the spherically symmetric
static metric can be written in the standard form \cite{c4}:

\begin{eqnarray}
ds^2=-e^\gamma dt^2+e^\alpha dr^2+e^\beta \left( d\theta ^2+\sin ^2\theta
\,d\varphi ^2\right) ,  \label{2.3}
\end{eqnarray}

In the case $\mu $= const standard Einstein fields equations yield

\begin{eqnarray}
\begin{array}{c}
g=\left( 
\begin{array}{cccc}
1 & 0 & 0 & 0 \\ 
0 & \left( \kappa ~\sinh \left( \frac r\kappa \right) \right) ^2 & 0 & 0 \\ 
0 & 0 & \left( \kappa ~\sinh \left( \frac r\kappa \right) \sin \theta
\right) ^2 & 0 \\ 
0 & 0 & 0 & c_2\left( 2c_1+\kappa \cosh \left( \frac r\kappa \right) \right)
^2
\end{array}
\right) \\ 
\\ 
p=-\frac{2c_1+3\kappa ~\cosh \left( \frac r\kappa \right) }{8\pi ~\kappa
^2~\left( 2c_1+\kappa ~\cosh \left( \frac r\kappa \right) \right) }
\end{array}
\label{12.2}
\end{eqnarray}

where c$_1$ and c$_2$ arbitrary constant and $\kappa =\sqrt{\frac 3{8\pi
~\mu }}.$ There is formal connection between the lower-dimensional theory
with a temporal scalar field and that of Einstein, but there are differences
and the physical interpretation is quite different. We should note that
despite the mathematical equivalence of the two solutions, that fact that
they come from different physical theories makes them conceptually distinct.
To conclude our discussion of the Lobachevsky geometry in theories of
gravity we make some remarks about Newtonian physics. Newtonian gravity in
Lobachevskyan space is defined by the fundamental solution of the Poisson
equation

\begin{equation}
\triangle \Phi =~4\pi G_N~\mu  \label{e2.8}
\end{equation}

where $\Phi $ is a gravitation potential acting on a test body and produced
by the mass $\mu $. In this note we focus on static spherically symmetric
constant density stars with metric

\begin{eqnarray}
ds^2=d\rho ^2+r^2\left( d\theta ^2+\sin ^2\theta \,d\varphi ^2\right) ,
\label{e2.34}
\end{eqnarray}

then the Laplace operator in the Lobachevskyan space is defined as:

\begin{eqnarray}
\triangle =\frac 1{r^2}\frac \partial {\partial \rho }r^2\frac \partial
{\partial \rho }+\frac 1{r^2}\left( \frac 1{\sin \theta }\frac \partial
{\partial \theta }\sin \theta \frac \partial {\partial \theta }+\frac 1{\sin
^2\theta \,}\frac{\partial ^2}{\partial \varphi ^2}\right) ,  \label{e2.33}
\end{eqnarray}

The difference between this and Euclidean geometries consists in the
dependence of r on $\rho $: in the Euclidean geometry r = $\rho $ while in
the Lobachevskyan one \cite{c6}

\begin{equation}
r=~\kappa ~\sinh \left( \frac \rho \kappa \right)  \label{e2.34}
\end{equation}

After simple manipulations, one can obtain the solution of equation (\ref
{e2.8})

\begin{equation}
\Phi =c_2~+\kappa ~\left( 2\pi G_N~\rho ~\mu +c_1\right) \coth \left( \frac
\rho \kappa \right)  \label{e2.35}
\end{equation}

with c1 and c2 a constants arising from integration.

\section{Conclusion}

During the recent years, there has been a lot of activity on theories with
extra dimension. The correspondence between extra and lower dimensions
theories is of potential importance in view of great deal of effort that is
currently going into study of lower dimensions theories, and particularly in
view of radical differences between such theories and 4- dimensional
gravity. Its interesting that the dependence of the metric on the extra
coordinates leads to, a new force term. Moreover, the particle will move
under influence of tow forces; the gravitational one and extra force which
does depend on the velocity \cite{c5}.

In the spirit of Newtonian theory, we separate the temporal scalar field
from the four-dimensional metric context of general relativity. In this
scheme space and time are treated in completely different ways. In contrast
with 2+1 dimension picture dynamics induced by the new temporal field
contains all features that one expects in the Einstein theory of gravitation.

The influence of the gravitational field upon the properties of solutions of
material field equations depends essentially on the matter distribution. The
simplest function of it used in gravitational theories is a constant
density. This approach gives interesting results, we obtain the solution of
field equations to be described by Lobachevsky geometry.

Static spherically symmetric perfect fluid models are also interest in
comparison with Newtonian theory. The replacement of Euclidian by
Lobachevsky geometry in Newtonian dynamics allows one the possibility to
explain the form of the rotation velocities of galaxies without the dark
matter hypothesis \cite{c6}. Comparing Newtons expression of the
gravitational force to expression of relativistic theory in Lobachevsky
space we find that the curvature of space depends from the energy density of
matter. Thus in this sense the temporal scalar field generates additional
effective sources and these in turn act to curve the 3-dimensional spacetime
too.

It is useful to illustrate the difference between the two theories, the
Newtonian and Relativistic ones, by using a Lobachevsky space as example.
This avenue has already proved fruitful in the case of low dimensional
configuration. 

\vspace*{1.5cm}

{\bf Acknowledgement. }Author wishes to thank prof. F.Kusmartsev for
fruitful discussion.

\end{document}